\documentclass[12pt]{article}

\usepackage{epsfig}
\usepackage{amssymb,amsmath}
\usepackage{graphicx, caption}
\usepackage{epstopdf}
\usepackage{color}
\usepackage{setspace}

\newcommand{\bea}{\begin{eqnarray}}
\newcommand{\eea}{\end{eqnarray}}

 \makeatletter
\def\alt{\mathrel{\mathpalette\gl@align<}}
\def\agt{\mathrel{\mathpalette\gl@align>}}
\def\gl@align#1#2{\lower.6ex\vbox{\baselineskip\z@skip\lineskip\z@
\ialign{$\m@th#1\hfil##\hfil$\crcr#2\crcr\sim\crcr}}} \makeatother

\begin{document}
%

\begin{center}
  \baselineskip 20pt {\Large\bf

No-Scale $\mu$-Term Hybrid Inflation

}
\vspace{1cm}

{\large Lina Wu$^{a,b,}$\footnote{E-mail:wulina@std.uestc.edu.cn},
 Shan Hu$^{c,}$\footnote{E-mail:hushan@itp.ac.cn},
Tianjun Li$^{a,b,d,}$\footnote{E-mail:tli@itp.ac.cn}}
\vspace{.5cm}

{\baselineskip 20pt \it

$^a$School of Physical Electronics, University of Electronic Science and Technology of China,
Chengdu 610054, P. R. China \\

\vspace{2mm}
$^b$Key Laboratory of Theoretical Physics and
Kavli Institute for Theoretical Physics China (KITPC),
Institute of Theoretical Physics, Chinese Academy of Sciences,
Beijing 100190, P. R. China \\

\vspace{2mm}
$^c$Department of Physics and Electronic Technology, Hubei University, Wuhan 430062, P. R. China \\

\vspace{2mm}
$^d$School of Physical Sciences, University of Chinese Academy of Sciences,
  Beijing 100049, P. R. China \\

}
\vspace{.5cm}

 {\bf Abstract}
\end{center}

To solve the fine-tuning problem in $\mu$-Term Hybrid Inflation,  we will realize
the supersymmetry scenario with the TeV-scale supersymmetric particles
and intermediate-scale gravitino from anomaly mediation, which can be consistent
with the WMAP and Planck experiments. Moreover,
we for the first time propose the $\mu$-term hybrid inflation
in no-scale supergravity.
With four Scenarios for the $SU(3)_C\times SU(2)_L\times SU(2)_R\times U(1)_{B-L}$ model,
we show that the correct scalar spectral index $n_s$
can be obtained, while the tensor-to-scalar ratio $r$ is prediced to be tiny,
about $10^{-10}-10^{-8}$. Also, the $SU(2)_R\times U(1)_{B-L}$ symmetry breaking scale
is around $10^{14}$~GeV, and all the supersymmetric particles except
gravitino are around TeV scale while gravitino mass is around $10^{9-10}$~GeV.
Considering the complete potential terms
linear in $S$, we for the first time show that the tadpole term,
which is the key for such kind of inflationary models to be consistent
with the observed scalar spectral index, vanishes after inflation. Thus,
to obtain the $\mu$ term, we need to generate the supersymmetry breaking
soft term  $A^{S \Phi \Phi'}_{\kappa} \kappa S \Phi \Phi'$ due to
$A^{S \Phi \Phi'}_{\kappa}=0 $ in no-scale supergravity, where $\Phi$ and $\Phi'$
are vector-like Higgs fields at high energy. We show that the proper
$A^{S \Phi \Phi'}_{\kappa} \kappa S \Phi \Phi'$ term can be obtained in the M-theory
inspired no-scale supergravity. We also point out that
$A^{S \Phi \Phi'}_{\kappa}$ around 700~GeV can be generated via
the renormalization group equation running from string scale.

\thispagestyle{empty}

\newpage

\addtocounter{page}{-1}

\baselineskip 18pt



It is well-known that our Universe may experience an accelerated expansion, {\it i.e.},
inflation~\cite{Staro, Guth:1980zm, Linde:1981mu, Albrecht:1982wi}, at a very early stage of
evolution, as suggested by the observed temperature
fluctuations in the cosmic microwave background radiation (CMB).
From the particle physics point of view, supersymmetry is the most promising
extension for the Standard Model (SM). In particular,
the scalar masses can be stabilized, and superpotential is
non-renormalized. Because gravity is also very important in the
early Universe, it seems to us that supergravity theory is
a natural framework for inflationary model building~\cite{SUGRA}.



The F-term  hybrid inflation in a supersymmetric high energy model with gauge
symmetry $G$ has a renormalizable superpotential $W$ and
a canonical K\"{a}hler potential $K$~\cite{DSS, EJC}. In particular,
the $Z_2$ $R$-parity in the Supersymmetric SMs (SSMs) is extended to a continuous $U(1)_R$ symmetry,
which determines superpotential. With the minimal $W$ and $K$,
the gauge symmetry $G$ is broken down to a subgroup $H$  at the end of inflation.
For the supersymmetric high energy model, in general, we can consider either
a left-right model with gauge symmetry
$SU(3)_C\times SU(2)_L\times SU(2)_R\times U(1)_{B-L}$, or
a Grand Unified Theory (GUT) such as  $SU(5)$ model, flipped $SU(5)\times U(1)_X$ model,
or Pati-Salam $SU(4)_c \times SU(2)_L \times SU(2)_R$ model~\cite{Pati-Salam}.
While $H$ can be the SM  or SM-like gauge group, etc.

In the original supersymmetric hybrid inflation
models, the quantum corrections arising from supersymmetry breaking drive inflation,
  and the scalar spectral index  was predicted to be $n_s = 1 - 1/N \simeq 0.98$~\cite{DSS},
  where $N= 60$ denotes the number of e-foldings necessary to resolve the horizon and flatness problems
  in Big Bang cosmology. Interestingly,  with  a class of linear supersymmetry breaking soft terms
  in the inflationary potential~\cite{RSW, buchm}, such kind of models can be highly consistent
with the observed scalar spectral index values
of $n_s=0.96-0.97$ from the WMAP~\cite{WMAP9y} and Planck satellite experiments~\cite{Planck2015}
as well. In particular, the corresponding supersymmetry breaking $A$-term for the linear superpotential term
can be around TeV scale~\cite{RSW, buchm}.

As we know, in the Minimal SSM (MSSM), there exists a well-known $\mu$ problem.
However, the $\mu H_d H_u$ term is forbidden by $U(1)_R$ symmetry, where
$H_u$ and $H_u$ are one pair of  Higgs fields in the SSMs. With the
linear supersymmetry breaking soft term  after inflation,
the inflaton field $S$ acquires a Vacuum Expectation Value (VEV). Thus, the
$\mu$ problem can be solved if there exists a superpotential term $\lambda S H_d H_u$,
as proposed by  Dvali, Lazarides and Shafi (DLS)~\cite{DLS}.
Assuming the minimal $K$, the magnitude of $\mu$ is typically around
the gravitino mass $m_G$~\cite{DLS}. Recently, such scenario has been studied
in details~\cite{Okada:2015vka}.
With the reheating and cosmological gravitino constraints,
it was found that a consistent inflationary scenario gives rather concrete predictions regarding
supersymmetric dark matter and Large Hadron Collider (LHC) phenomenology.
Especially, the gravitino must be sufficiently heavy ($m_G \gtrsim 5 \times 10^7$ GeV)
so that it decays before the freeze out of the lightest supersymmetric particle (LSP) neutralino,
which is the dark matter candidate. Moreover, the wino with mass $\simeq 2$ TeV becomes
a compelling dark matter candidate. And
the supersymmetry breaking scalar mass  $M_0$ is expected to be of the same order as $m_G$ or larger,
   which can reproduce a SM-like Higgs boson mass $\simeq 125$ GeV for suitable $\tan \beta$ values,
   where $\tan \beta$ is the ratio of the VEVs for  $H_u$ and $H_d$.
Depending on the underlying gauge symmetry $G$ associated with inflationary scenario,
   the observed baryon asymmetry in the Universe can be explained via leptogenesis \cite{Fuku-Yana, LS}.
The compelling examples of $G$, in which the DLS mechanism can be successfully merged with inflation, contain
   $U(1)_{B-L}$, $SU(2)_L \times SU(2)_R \times U(1)_{B-L}$, and flipped $SU(5)\times U(1)_X$.
The other examples of $G$ are $SU(5)$ and $SU(4)_C \times SU(2)_L \times SU(2)_R$~\cite{Pati-Salam},
but there may exist monopole problem.

In short, in the recent study~\cite{Okada:2015vka}, to solve the gravitino problem in
the $\mu$-term hybrid inflation, Okada and Shafi showed that
the sfermions, Higgsinos, and gravitino are heavy around $10^7$~GeV while
gauginos are light around TeV, which are similar to the split supersymmetry~\cite{splitSUSY, Haba:2005ux}\footnote{The supersymmetric hybrid inflation model with a no-scale form of the Kahler potential, which is based on a Heisenberg symmetry, has been studied before to solve the $\eta$ problem \cite{1,2}.
}. Thus,
the supersymmetry solution to gauge hierarchy problem is at least partly gone, {\it i.e.},
there exists big fine-tuning around $10^{-10}$. On the other hand,
even if the corresponding supersymmetry breaking $A$-term for the linear superpotential term
is around TeV scale~\cite{RSW, buchm}, we can still obtain the observed scalar spectral index values
of $n_s=0.96-0.97$ from the WMAP~\cite{WMAP9y} and Planck satellite experiments~\cite{Planck2015}.
Therefore, to solve this problem, we do need the
  supersymmetry scenario, which can have the TeV-scale supersymmetric particles
  (sparticles) in the SSMs while intermediate-scale heavy gravitino.
The well-known example is no-scale supergravity~\cite{Cremmer:1983bf}
or its generalization. In this paper, we shall realize such supersymmetry scenario via
anomaly mediation~\cite{Haba:2005ux}. In addition,
we for the first time propose the $\mu$-term hybrid inflation 
in no-scale supergravity\footnote{The gravitino mass can be around the TeV scale if there exists extra D-term contribution \cite{3}.}. We discuss it in details, and find
some interesting results different from the previous study on the $\mu$-term hybrid inflation.

First, with anomaly mediation, we will derive the supersymmetry scenario, where
the sparticles are light while gravitino is heavy~\cite{Haba:2005ux}.
We consider the K\"ahler potential and superpotential as follows
\bea
K &=& -3 M_{Pl}^2 \left(z+{\bar z}+\epsilon f(z, {\bar z})\right) {\bar X} X +\sum_Y {\bar Y} Y ~,~\,
\eea
\bea
W &=& X^3 W_0 + S \left(\kappa X^2 M^2 - \kappa \Phi' \Phi + \lambda H_d H_u \right)~,~\,
\eea
where $M_{Pl}$ is the reduced Planck scale, $z$ and $X$ are respectively a hidden sector superfield
and a compensator multiplet ($X=1+F_X$), $Y$ denotes all the other superfields, 
$\epsilon$ is a small parameter, $W_0$ is a constant
superpotential, and $\Phi'$ and $\Phi$ are the Higgs fields which breaks the high-scale gauge symmetry
in the F-term hybrid inflation~\cite{DSS, EJC}. Similar to the no-scale supergravity,
the scalar potential vanishes in the limit $\epsilon \rightarrow 0$. Considering the
equations of motion for the auxiliary fields, we obtain
\bea
F_X \simeq -\frac{W_0^{\dagger}}{M_{Pl}^2} \epsilon  f_{{\bar z}z} = - \epsilon m_G f_{{\bar z}z}~,~~~
F_z \simeq \frac{W_0^{\dagger}}{M_{Pl}^2}~=~m_G~,~
\eea
for small $\epsilon$. Here, we define
$f_{{\bar z}z} \equiv \partial^2 f(z, {\bar z})/ \partial {\bar z}  \partial z$,
and $m_G$ is gravitino mass. So the scalar potential
becomes
\bea
V =-3 F_X W_0 \simeq 3\frac{|W_0|^2}{M_{Pl}^2} \epsilon  f_{{\bar z}z} = 3 \epsilon m^2_G M_{Pl}^2 f_{{\bar z}z}~.~
\eea
For example, assuming $f_{{\bar z}z} = (|z|^2-1/4)^2-1$, we get the minimum for the scalar potential
at $\langle z\rangle =1/2$
\bea
V_{\rm min} \simeq  -3 \epsilon m^2_G M_{Pl}^2 ~,~
\label{Potential-Vacuum}
\eea
which is an AdS vacuum. 
Thus, we have $F_X\simeq \epsilon m_G << m_G$. Because the supersymmetry breaking soft terms in
the SSMs are proportional to $F_X$ via anomaly mediation, we obtain the supersymmetry breaking scenario
which has TeV-scale sparticles and intermediate-scale gravitino.
In particular, the supersymmetry breaking linear term for $S$  is given by
\bea
V &=& -4 \kappa F_X M^2 S + {\rm H.C} \simeq -4 \kappa \epsilon m_G M^2 S + {\rm H.C}~.~\,
\eea
From the numerical studies in Refs.~\cite{RSW, buchm},
we can still obtain  the observed scalar spectral index values
of $n_s=0.96-0.97$ from the WMAP~\cite{WMAP9y} and Planck satellite experiments~\cite{Planck2015}
as well. By the way, the AdS vacuum given by Eq.~(\ref{Potential-Vacuum}) can be lifted to the
Minkowski vacuum by considering the $F$-term and $D$-term contributions in the anomalous
$U(1)$ theory inspired from string models~\cite{Haba:2005ux}.

In the following, we shall embed the previous $\mu$-term hybrid inflation scenario
into no-scale supergravity framework,
{\it i.e.}, we propose the $\mu$-term hybrid inflation in no-scale supergravity
where $\mu$ term is generated via the VEV of inflaton field after inflation.
We introduce  a conjugate pair of vector-like Higgs fields $\Phi$ and $\Phi'$,
which breaks $G$ down to the SM or SM-like gauge symmetry.
Considering four Scenarios for the $SU(3)_C\times SU(2)_L\times SU(2)_R\times U(1)_{B-L}$ model,
we show that the correct scalar spectral index $n_s$
can be obtained, while the tensor-to-scalar ratio $r$ is prediced to be
tiny, about $10^{-10}-10^{-8}$. Also,  the $SU(2)_R\times U(1)_{B-L}$ symmetry breaking scale
is around $10^{14}$~GeV, and all the supersymmetric particles except
gravitino are around TeV scale while gravitino mass is around $10^{9-10}$~GeV.
We present the complete potential terms that are linear in $S$, and for the first time
we show that the tadpole term, which is the key for such kind of inflationary models to be consistent
with the observed scalar spectral index, vanishes after inflation or say gauge symmetry
$G$ breaking. Thus, to reproduce the $\mu$ term, we need to generate
the supersymmetry breaking soft term  $A^{S \Phi \Phi'}_{\kappa} \kappa S \Phi \Phi'$
since  we have $A^{S \Phi \Phi'}_{\kappa}=0 $ in no-scale supergravity.
We show that the supersymmetry breaking soft term  $A^{S \Phi \Phi'}_{\kappa} \kappa S \Phi \Phi'$
can be generated properly in the M-theory inspired no-scale supergravity which has no-scale
supergravity at the leading
or lowest order~\cite{Horava:1996ma, Li:1997sk, Lukas:1997fg, Nilles:1998sx, Li:1998rn}.
We also point out that the $A^{S \Phi \Phi'}_{\kappa} \kappa S \Phi \Phi'$
term with $A^{S \Phi \Phi'}_{\kappa}$ around 700~GeV can be obtained
via the renormalization group equation (RGE) running
from string scale~\cite{Ellis:2001kg, Schmaltz:2000gy, Ellis:2010jb, Li:2010ws}.
Therefore, we solve the fine-tuning problem in the previous $\mu$-term hybrid
inflation, and propose the no-scale $\mu$-term hybrid
inflation models where the sparticles in the SSMs are around TeV scale while
gravitino is around $10^{9-10}$~GeV.

Let us present our model in the following. The K\"ahler potential is
\bea
K &=& \overline{S} S - 3 {\rm ln}\left(T+\overline{T}-2\overline{C_i} C_i \right)~,~\,
\eea
where $T$ is a modulus, and $C_i$ are matter/Higgs fields in the supersymmetric SMs
which include $\Phi$, $\Phi'$, $H_u$, and $H_d$. To simiplify the discussions,
we will assume $\langle T \rangle =1/2$ in the following study.

Assuming $S$ and superpotential have charge 2 while  $\Phi$, $\Phi'$,
$H_u$ and $H_d$ are neutral under the $U(1)_R$ R-symmetry, we obtain the $U(1)_R$ invariant
inflaton superpotential~\cite{DLS}
\begin{equation}
     W = S \left( \kappa \Phi' \Phi - \kappa M^2 + \lambda H_d H_u \right)~.\,
\end{equation}
To realize the correct symmetry breaking pattern after inflation, we require
$\lambda > \kappa$~\cite{DLS}. In particular, the $\mu H_d H_u$ term is forbidden
by the $U(1)_R$ R-symmetry, and then such term can be generated only after
$U(1)_R$ R-symmetry is broken down to a $Z_2$ symmetry, for example,
by the VEV of $S$.

Assuming that the F-term of $T$ breaks supersymmetry, we obtain the following scalar potential
which is linear in $S$
\bea
V &\supset & m_G S \left( \kappa \Phi' \Phi - \kappa M^2 + \lambda H_d H_u \right)+{\rm H.C.}~.~\,
\label{LN-SP}
\eea
 As a side remark,
for Polonyi model, we will have an extra ($-2$) factor in
the above tadpole term due to the $-3|W|^2$ contribution.
During inflation, we have $\langle \Phi \rangle = \langle \Phi' \rangle =0$,
as well as a tadpole term for $S$
\bea
V &\supset & - \kappa m_G M^2 S + {\rm H.C.} ~.~\,
\eea
After inflation (or say after gauge symmetry $G$ breaking)
and neglecting the VEVs of $H_u$ and $H_d$, we have $\langle \Phi \rangle = \langle \Phi' \rangle =M$,
and then the above tadpole term vanishes.
To obtain the $\mu$ term which is forbidden by
$U(1)_R$ symmetry, we need to generate the
tadpole term of $S$, which will be discussed below.

With the supersymmetry breaking soft mass term as well as
the radiative  and supergravity corrections, we obtain
the inflationary potential as follows
\bea
V(\phi) &=& \frac{1}{2} m_\phi^2 \phi^2 +
m^4 \left( 1+ \alpha \ln \left[\frac{\phi}{\phi_0}\right] + \frac{3 \phi^2}{2M^2_{\rm Pl}}
+ \frac{7\phi^4}{8M^4_{\rm Pl}}\right) -  \sqrt{2} m_G m^2 \phi~,~
 \label{Vinf}
 \eea
 where  $m = \sqrt{\kappa} M$, $\phi$ is the real part of $S$,
 $m_\phi$ is the supersymmetry breaking soft mass, $M_{\rm Pl}$ is the reduced Planck scale,
the renormalization scale ($Q$) is chosen to be equal
 to the initial inflaton VEV $\phi_0$, and the coefficient $\alpha \ll1$ is given by
\begin{equation}
\alpha =  \frac{1}{4 \pi^2} \left( \lambda^2 + \frac{N_{\Phi}}{2} \kappa^2 \right)~.~\,
\end{equation}
In particular, the negative sign of the linear term is essential to generate the correct value for
the spectral index. Without this linear term, the scalar spectral index $n_s$
is predicted to lie close to $0.98$, as shown in Ref.~\cite{DSS}.
Moreover, both $\phi^2$ and $\phi^4$ terms arise from the leading supergravity contribution,
the quadratic  supersymmetry breaking soft term can be ignored relative to
the liner term in Eq.~(\ref{Vinf})~\cite{RSW, buchm},
and the imaginary part of $S$ is assumed to
stay constant during inflation
[For a more complete discussion of this last point, see Ref.~\cite{buchm}.].
Thus, the inflaton potential can be simplified to
\bea
V(\phi) &=&
m^4 \left( 1+ \alpha \ln \left[\frac{\phi}{\phi_0}\right] + \frac{3 \phi^2}{2M^2_{\rm Pl}}
+ \frac{7\phi^4}{8M^4_{\rm Pl}}\right) -  \sqrt{2} m_G m^2 \phi~,~
 \label{Vinflation}
 \eea

In the following discussions, to be concrete, we consider the left-right model with gauge symmetry
$SU(3)_C\times SU(2)_L\times SU(2)_R\times U(1)_{B-L}$. Because
$\Phi$ and $\Phi'$  respectively
have quantum numbers $(\mathbf{1}, \mathbf{1},  \mathbf{2},  \mathbf{1/2})$
and  $(\mathbf{1}, \mathbf{1},  \mathbf{2},  \mathbf{-1/2})$, we get $N_{\Phi}=2$.
For simplicity, we set $\gamma\equiv \lambda/\kappa =2$,
and then have $\tilde{\gamma} \equiv \sqrt{\gamma^2+N_{\Phi}/2} =  \sqrt{5}$. Moreover, 
it seems to us that the $\phi^4$ term can be neglected as well. Therefore,
we will study the following four scenarios where
the power spectrum $\Delta_R^2=2.20 \times 10^{-9}$ from the Planck 2015 results~\cite{Planck2015}
has been explained simultaneously: \\

{\bf Scenario I.} The potential in Eq.~(\ref{Vinf}) with $m_\phi \simeq m_G$. \\

\begin{figure}[htb]
\centering
\includegraphics[height=8cm]{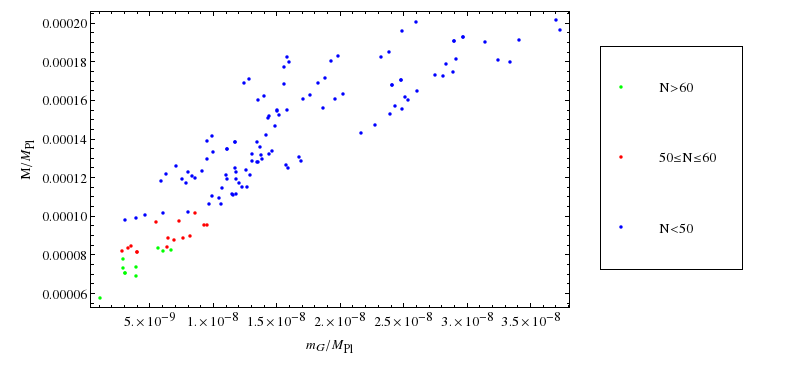}
\caption {The allowed numerical values for $M$ and $m_{G}$ to get $0.955 \leq n_s \leq 0.977$ and $27 \leq N \leq 72$
  for the potential in Eq.~(\ref{Vinf}) with $m_\phi \simeq m_G$.
  Here, we have $0.358747\leq \kappa \leq 1.02244$.} \label{Pot5 with phi4}
\end{figure}

To obtain $0.955 \leq n_s \leq 0.977$ within about $1\sigma$ range of the Planck 2015 results~\cite{Planck2015}
and the e-folding number $27 \leq N \leq 72$, we present the numerical values of $M$ and $m_G$
for the viable points in Fig.~(\ref{Pot5 with phi4}), which are normalized
by the reduced Planck scale $M_{Pl}=2.43 \times 10^{18}~{\rm GeV}$.
 The corresponding range of $\kappa$ is
$0.358747\leq \kappa \leq 1.02244$. According to the figure, we can see with the increasing of $N$,
both $M$ and $m_{G}$ decrease. The best fit point with the Planck results has
$n_{s}=0.964677$, $r=1.32516\times 10^{-9}$, and $N=55$, which can be obtained
by choosing $\kappa=0.46682$, $M=1.19883 \times 10^{-4}M_{Pl}\approx 2.913\times 10^{14}~{\rm GeV}$, and 
$m_{G}=2.8227 \times 10^{-9} M_{Pl}\approx 6.859\times10^{9}~{\rm GeV}$. Moreover,
the minimal value of $M$ locates at $M= 7.44428 \times 10^{-5}M_{Pl}\approx 1.80896\times10^{14}~{\rm GeV}$
with the corresponding $\kappa=0.855067$ and $m_ G= 3.96719\times10^{-9} M_{Pl} \approx 9.64027\times10^{9}~{\rm GeV}$.
The corresponding inflationary observables and number of e-folding are
$n_{s}=0.966343$, $r=5.94759\times10^{-10}$, and $N=71$, respectively. 
Also, the minimal value of $m_G$ locates at $m_G=1.09362\times10^{-9}M_{Pl} \approx 2.6575\times10^{9}~{\rm GeV}$
with the corresponding $\kappa=0.371718$ and $M=9.42889 \times 10^{-5}M_{Pl}\approx 2.29122\times10^{14}~{\rm GeV}$.
The corresponding inflationary observables and number of e-folding are
$n_{s}=0.958972$, $r=3.2656\times10^{-10}$, and $N=66$, respectively. \\

{\bf Scenario II.} The potential in Eq.~(\ref{Vinf}) with $m_\phi \simeq m_G$ and without the $\phi^4$ term.\\

\begin{figure}[htb]
\centering
\includegraphics[height=8cm]{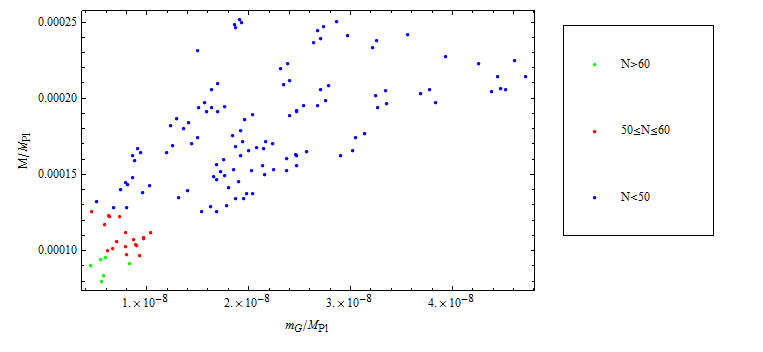}
\caption {The allowed numerical values for $M$ and $m_{G}$ to get $0.955 \leq n_s \leq 0.977$ and $27 \leq N \leq 72$ for the potential in Eq.~(\ref{Vinf}) with $m_\phi \simeq m_G$ and without the $\phi^4$ term.
Here, we have $0.554916\leq \kappa \leq 1.03625$.} \label{Pot5 without phi4}
\end{figure}

To obtain $0.955 \leq n_s \leq 0.977$ within about $1\sigma$ range of the Planck 2015 results
and the e-folding number $27 \leq N \leq 72$, we present the numerical values of $M$ and $m_G$
for the viable points in Fig.~(\ref{Pot5 without phi4}).  The corresponding range of $\kappa$ is
$0.554916\leq \kappa \leq 1.03625$. The best fit point with the Planck results has
$n_{s}=0.964383$, $r=4.99194\times 10^{-8}$, and $N=54$, which can be obtained
by taking $\kappa=0.694001$, $M=2.46883 \times 10^{-4}M_{Pl}\approx 5.999\times 10^{14}~{\rm GeV}$,
and $m_{G}=2.73085 \times 10^{-8} M_{Pl}\approx 6.636\times10^{10}~{\rm GeV}$.
In addition, 
the minimal value of $M$ locates at $M= 7.95428 \times 10^{-5}M_{Pl}\approx 1.93289\times10^{14}~{\rm GeV}$ with the corresponding $\kappa=0.94457$ and $m_G= 5.59342\times10^{-9} M_{Pl} \approx 1.3592\times10^{10}~{\rm GeV}$.
The corresponding inflationary observables and number of e-folding are
$n_{s}=0.962104$, $r=9.10999\times10^{-10}$, and $N=67$, respectively. 
Also, the minimal value of $m_G$ locates at $m_G=4.52267\times10^{-9}M_{Pl} \approx 1.09901\times10^{10}~{\rm GeV}$ with the corresponding $\kappa=0.764382$ and $M=9.05332 \times 10^{-5}M_{Pl}\approx 2.19996\times10^{14}~{\rm GeV}$.
The corresponding inflationary observables and number of e-folding are
$n_{s}=0.959513$, $r=1.0716\times10^{-9}$, and $N=64$, respectively. \\

{\bf Scenario III.}  The potential in Eq.~(\ref{Vinflation}). \\

\begin{figure}[htp]
\centering
\includegraphics[height=8cm]{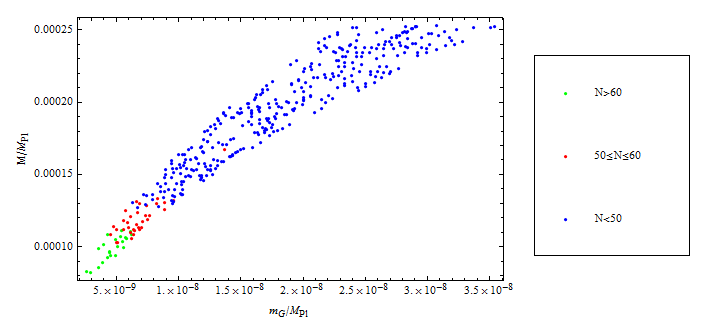}
\caption {The allowed numerical values for $M$ and $m_{G}$ to get $0.955 \leq n_s \leq 0.977$ and $27 \leq N \leq 72$ for the potential in Eq.~(\ref{Vinflation}). Here, we have $0.643221\leq \kappa \leq 0.799225$.} \label{Pot7 with phi4}
\end{figure}

To obtain $0.955 \leq n_s \leq 0.977$ within about $1\sigma$ range of the Planck 2015 results and
the e-folding number $27 \leq N \leq 72$, we present the numerical values of $M$ and $m_G$
for the viable points in Fig.~(\ref{Pot7 with phi4}). The corresponding range of $\kappa$ is
$0.643221\leq \kappa \leq 0.799225$. The best fit point with the Planck results has $n_{s}=0.965618$,
$r=3.52383\times10^{-9}$, and $N=53$, which can be obtained by choosing
$\kappa=0.76438$, $M=1.2187 \times 10^{-4}M_{Pl}\approx2.96139\times10^{14}~{\rm GeV}$, and
$m_G=7.66695\times10^{-9} M_{Pl}\approx1.86307\times10^{10}~{\rm GeV}$. Moreover,
the minimal value of $M$ locates at $M= 8.20741 \times 10^{-5}M_{Pl}\approx 1.9944\times10^{14}~{\rm GeV}$ with
the corresponding $\kappa=0.698536$ and $m_G= 2.89804\times10^{-9} M_{Pl} \approx 7.04225\times10^{9}~{\rm GeV}$.
The corresponding inflationary observables and number of e-folding are
$n_{s}=0.955277$, $r=6.17585\times10^{-10}$, and $N=70$, respectively. 
Also, the minimal value of $m_G$ locates at $m_G=2.56278\times10^{-9}M_{Pl} \approx 6.22754\times10^{9}~{\rm GeV}$ with the corresponding $\kappa=0.654176$ and $M=8.24695 \times 10^{-5}M_{Pl}\approx 2.00401\times10^{14}~{\rm GeV}$.
The corresponding inflationary observables and number of e-folding are
$n_{s}=0.958115$, $r=5.58929\times10^{-10}$, and $N=71$, respectively. \\

{\bf Scenario IV.} The potential in Eq.~(\ref{Vinflation}) without the $\phi^4$ term. \\

\begin{figure}[htp]
\centering
\includegraphics[height=8cm]{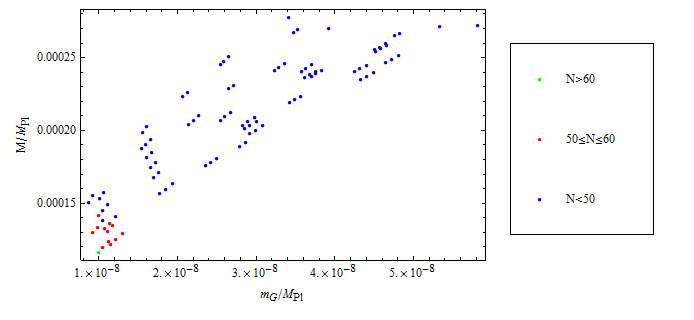}
\caption {The allowed numerical values for $M$ and $m_{G}$ to get $0.955 \leq n_s \leq 0.977$ and $27 \leq N \leq 72$ for the potential in Eq.~(\ref{Vinflation}) without the $\phi^4$ term. Here, we have
  $0.664356 \leq \kappa \leq 0.948324$.} \label{Pot7 without phi4}
\end{figure}

To obtain $0.955 \leq n_s \leq 0.977$ within about $1\sigma$ range of the Planck 2015 results and
the e-folding number $27 \leq N \leq 72$, we present the numerical values of $M$ and $m_G$
for the viable points in Fig.~(\ref{Pot7 with phi4}).  The corresponding range of $\kappa$ is
$0.664356 \leq \kappa \leq 0.948324$. The best fit point with the Planck results
has $n_{s}=0.966235$, $r=5.4042\times10^{-9}$, and $N=52$, which can be obtained
by taking $\kappa=0.79576$, $M=1.33317 \times 10^{-4}M_{Pl}\approx3.23961\times10^{14}~{\rm GeV}$,
and $m_G=9.81092\times10^{-9} M_{Pl}\approx2.38405\times10^{10}~{\rm GeV}$. In addition,
the minimal value of $M$ locates at $M= 1.16234 \times 10^{-4} M_{Pl}\approx 2.8245\times10^{14}~{\rm GeV}$ with
the corresponding $\kappa=0.917431$ and $m_G= 9.91267\times10^{-9}  M_{Pl} \approx 2.40878\times10^{10}~{\rm GeV}$.
The corresponding inflationary observables and number of e-folding are
$n_{s}=0.976999$, $r=3.96363\times10^{-9}$, and $N=62$, respectively.
Also, the minimal value of $m_G$ locates at $m_G=8.71055\times10^{-9} M_{Pl}\approx 2.11666\times10^{10}~{\rm GeV}$
with the corresponding $\kappa=0.664356$ and $M=1.50466 \times 10^{-4}M_{Pl}\approx 3.65632\times10^{14}~{\rm GeV}$.
The corresponding inflationary observables and number of e-folding are
$n_{s}=0.970739$, $r=6.36454\times10^{-9}$, and $N=48$, respectively. 

In short, from the above numerical studies, we find that  the observed scalar spectral index $n_s$
can be realized, but the tensor-to-scalar ratio $r$ is prediced to be tiny,
about $10^{-10}-10^{-8}$. Also, the $SU(2)_R\times U(1)_{B-L}$ symmetry breaking scale
is around $10^{14}$~GeV, and the gravitino mass is around $10^{9-10}$~GeV.
Thus, we do need the no-scale supergravity to realize the light sparticle spectrum.

Because gravitino is heavy and then unstable, we encounter the cosmological
gravitino problem~\cite{gravitino_problem}, which originates from the gravitino lifetime
\bea \label{taug}
 \tau_G \simeq 10^{4} \; {\rm sec} \times \left(  \frac{1\; {\rm TeV}}{m_G}\right)^3~.
\eea
To avoid the constraint on the neutralino abundance from gravitino decay,
  we assume that the LSP neutralino is still in thermal equilibrium when gravitino decays.
So the LSP neutralino abundance is not related to the gravitino yield.
Using a typical value of the ratio $x_F\equiv m_{{\tilde \chi}^0}/T_F \simeq 20$,
  where $T_F$ is the freeze out temperature of the LSP neutralino, this occurs for the gravitino lifetime
\bea
  \tau_G \lesssim 4 \times 10^{-10} \left( \frac{1 \text{ TeV}}{m_{{\tilde \chi}^0}} \right)^2.
\eea
Combining this with Eq.(\ref{taug}), we find
\bea
  m_G \gtrsim 4.6 \times 10^7 \text{ GeV} \left( \frac{m_{{\tilde \chi}^0}}{2 \text{ TeV}} \right)^{2/3}.
\label{LB_mG}
\eea
Therefore, such cosmological scenario favors a gravitino mass at an intermediate scale above $10^7$~GeV,
and the gravitino mass in our model satisfies this bound clearly.

Furthermore, after $SU(2)_R\times U(1)_{B-L}$ gauge symmetry breaking,
the leading tadpole term for $S$ in Eq.~(\ref{LN-SP}) vanishes.
Thus, to obtain the $\mu$ term which is forbidden
by $U(1)_R$ symmetry,
we need to generate the supersymmetry breaking soft
term  $A^{S \Phi \Phi'}_{\kappa} \kappa S \Phi \Phi'$. With it,
we get the VEV of $S$ as below
\begin{eqnarray}
  \langle S \rangle &=& \frac{A^{S \Phi \Phi'}_{\kappa}}{2\kappa}~.~\,
\end{eqnarray}
And then the $\mu$ term is given by
\begin{eqnarray}
 \mu  &=& \frac{\lambda}{2\kappa}A^{S \Phi \Phi'}_{\kappa}~.~\,
\end{eqnarray}
For $\lambda =2 \kappa$, we have
\begin{eqnarray}
 \mu  &=& A^{S \Phi \Phi'}_{\kappa}~.~\,
\end{eqnarray}

However, in no-scale supergravity, we have $A^{S \Phi \Phi'}_{\kappa}=0 $.
To solve this problem,
first,  we  consider M-theory on $S^1/Z_2$~\cite{Horava:1996ma}. For the
standard Calabi-Yau compactification at the leading order or lowest order, we can realize
no-scale supergravity~\cite{Li:1997sk}, and there exists
the next to leading order corrections~\cite{Lukas:1997fg, Nilles:1998sx, Li:1998rn}.
In particular, we can have the non-zero supersymmetry breaking
soft term $A^{S \Phi \Phi'}_{\kappa} \kappa S \Phi \Phi'$. To compare with no-scale supergravity, we consider
moduli dominant supersymmetry breaking, whose the supersymmetry breaking soft terms for universal
gaugino mass, scalar mass and trilinear soft term are~\cite{Li:1998rn}
\begin{eqnarray}
M_{1/2}&=&  \frac{x}{1+x} m_G ~,~ \,
\end{eqnarray}
\begin{eqnarray}
M_0&=& \frac{x}{3+x} m_G  ~,~\,
\end{eqnarray}
\begin{eqnarray}
A&=&- \frac{3x}{3+x} m_G  ~,~\,
\end{eqnarray}
where $0< x < 1$. For $x \sim 10^{-6}-10^{-7}$, we can indeed have the TeV-scale supersymmetry breaking
soft terms in the SSMs while gravitino mass is around $10^{9-10}$ GeV. Of course, there exists
some fine-tuning for $x$.

Another way to generate the $A^{S \Phi \Phi'}_{\kappa} \kappa S \Phi \Phi'$ term is from the
RGE running in no-scale supergravity~\cite{Ellis:2001kg, Schmaltz:2000gy, Ellis:2010jb, Li:2010ws}.
Because of $A=0$ from the no-scale boundary condition, 
we can neglect the Yukawa contributions and the RGE for  $A^{S \Phi \Phi'}_{\kappa} $ is
\begin{eqnarray}
  16\pi^2 \frac{dA^{S \Phi \Phi'}_{\kappa}}{dt} &=& -2\left(g_{B-L}^2 M_{B-L}+3 g_{2R}^2 M_{2R}\right)
\end{eqnarray}
before the $SU(2)_R\times U(1)_{B-L}$ gauge symmetry breaking, and
\begin{eqnarray}
  16\pi^2 \frac{dA^{S \Phi \Phi'}_{\kappa}}{dt} &=& -4 g_{1}^2 M_{1}
\end{eqnarray}
after the $SU(2)_R\times U(1)_{B-L}$ gauge symmetry breaking. Here, $t={\rm ln} \mu$, $g_{B-L}$, $g_{2R}$,
and $g_1$ are respectively gauge couplings for $U(1)_{B-L}$, $SU(2)_R$, and $U(1)_Y$,
and $M_{B-L}$, $ M_{2R}$, and $M_{1} $ are the corresponding gaugino masses.
The boundary condition for the gauge couplings at the $SU(2)_R\times U(1)_{B-L}$ gauge
symmetry breaking scale is
\begin{eqnarray}
  \frac{1}{g_1^2} &=& \frac{1}{g_{B-L}^2} + \frac{1}{g_{2R}^2}~.~\,
\end{eqnarray}
Because we do not present a complete model here, let us consider the simple case.
For no-scale supergravity, we should run the RGEs from the string scale, otherwise,
light stau will be the LSP~\cite{Ellis:2001kg, Schmaltz:2000gy, Ellis:2010jb, Li:2010ws}.
Thus, we run the RGE from string scale to the scale
around the masses of $S$, $\Phi$, and $\Phi'$. For $g_{B-L}=g_{2R}=1$ and
$M_{B-L}= M_{2R}=2$~TeV, assuming the constant gauge couplings and gaugino masses,
we get $\mu = A^{S \Phi \Phi'}_{\kappa}\simeq -700$~GeV for order one $\kappa$.
Of course, in such kind of the left-right models,
we generically need to introduce more particles, and the complete RGE study is
much more complicated. Note that if we have more particles above the
$SU(2)_R\times U(1)_{B-L}$ symmetry breaking scale, 
their gauge couplings will become larger at higher scale and then
the magnitude of  $A^{S \Phi \Phi'}_{\kappa}$ will be larger, which can give us larger $\mu$ term
if we want.
Therefore, we can indeed obtain the SSMs with TeV-scale supersymmetry
and the intermediate-scale heavy gravitino.

In summary, to solve the problem in the $\mu$-term hybrid inflation with canonical K\"ahler
potential,  we obtained the supersymmetry scenario which has the TeV-scale supersymmetric particles
and intermediate-scale gravitino from anomaly mediation. Moreover,
we for the first time proposed the $\mu$-term hybrid inflation in no-scale supergravity
where $\mu$ term is generated via the VEV of inflaton field after inflation.
Considering four Scenarios for the $SU(3)_C\times SU(2)_L\times SU(2)_R\times U(1)_{B-L}$ model,
we showed that the correct scalar spectral index $n_s$
can be obtained, while the tensor-to-scalar ratio $r$ is prediced to be
tiny, about $10^{-10}-10^{-8}$. Also, the $SU(2)_R\times U(1)_{B-L}$ symmetry breaking scale
is around $10^{14}$~GeV, and all the supersymmetric particles except
gravitino are around TeV scale while gravitino mass is around $10^{9-10}$~GeV.
With the complete potential terms linear in $S$, we for the first time
showed that the tadpole term, which is the key for such kind of inflationary models to be consistent
with the observed scalar spectral index, vanishes after inflation or say gauge symmetry
$G$ breaking. Thus, to obtain the $\mu$ term, we need to generate
the supersymmetry breaking soft term  $A^{S \Phi \Phi'}_{\kappa} \kappa S \Phi \Phi'$,
since  we have $A^{S \Phi \Phi'}_{\kappa}=0 $ in no-scale supergravity.
We showed that the supersymmetry breaking soft term  $A^{S \Phi \Phi'}_{\kappa} \kappa S \Phi \Phi'$
can be realized properly in the M-theory inspired no-scale supergravity which has no-scale
supergravity at the leading or lowest order. Also, we pointed out that
the $A^{S \Phi \Phi'}_{\kappa} \kappa S \Phi \Phi'$ term with $A^{S \Phi \Phi'}_{\kappa}$
around a few hundred GeVs can be reproduced
via the RGE running
from string scale. Therefore, we proposed the no-scale $\mu$-term hybrid
inflation models where the sparticles in the SSMs are around TeV scale while
gravitino is around $10^{9-10}$~GeV.

\section*{Acknowledgments}

We would like to thank Qaisar Shafi very much for suggesting the project,
and thank Qaisar Shafi and Nobuchika Okada very much for collaboration in
the initial stage of the project and helpful discussions.
This research was supported in part by the Natural Science Foundation of China under
grant numbers 11135003, 11275246, and 11475238 (T.L.).


\end{document}